\documentclass[12pt]{iopart}
\expandafter\let\csname equation*\endcsname\relax
\expandafter\let\csname endequation*\endcsname\relax
\usepackage{graphicx}
\usepackage{siunitx}
\usepackage[version=4]{mhchem}
\usepackage{hyperref} 
\usepackage[capitalize]{cleveref}
\usepackage[square,numbers]{natbib}
\usepackage{iopams}
\usepackage{caption}
\usepackage{subcaption}
\usepackage{multirow}
\usepackage{booktabs}
\usepackage[title]{appendix}

\newtheorem{definition}{Definition}%

\begin{document}

\title[Dis-GEN: Disordered crystal structure generation]{Dis-GEN: Disordered crystal structure generation}

\author{Martin Hoffmann Petersen$^1$*, Ruiming Zhu$^2$, Haiwen Dai$^{2,3}$, Savyasanchi Aggarwal$^{3,4}$, Nong Wei$^2$, Andy Paul Chen$^2$, Arghya Bhowmik$^1$, Juan Maria Garcia Lastra$^1$ and Kedar Hippalgaonkar$^{2,3}$*}

\address{1 Technical University of Denmark, Department of Energy Conversion and Storage, Lyngby, 2800, Denmark}
\address{2 School of Materials Science and Engineering, Nanyang Technological University, 50 Nanyang Avenue, Singapore, 639798, Singapore}
\address{3 Institute of Materials Research and Engineering, A*STAR, 2 Fusionopolis Way, Singapore, 138634, Singapore}
\address{4 Department of Chemistry, University College London, 20 Gordon St, London, WC1H 0AJ, United Kingdom}

\ead{mahpe@dtu.dk,kedar@ntu.edu.sg}

\begin{abstract}
A wide range of synthesized crystalline inorganic materials exhibit compositional disorder, where multiple atomic species partially occupy the same crystallographic site. As a result, the physical and chemical properties of such materials are dependent on how the atomic species are distributed among the corresponding symmetrical sites, making them exceptionally challenging to model using computational methods. 
For this reason, existing generative models cannot handle the complexities of disordered inorganic crystals. To address this gap, we introduce Dis-GEN, a generative model based on an empirical equivariant representation, derived from theoretical crystallography methodology. Dis-GEN is capable of generating symmetry-consistent structures that accommodate both compositional disorder and vacancies. The model is uniquely trained on experimental structures from the Inorganic Crystal Structure Database (ICSD) - the world's largest database of identified inorganic crystal structures. We demonstrate that Dis-GEN can effectively generate disordered inorganic materials while preserving crystallographic symmetry throughout the generation process. This approach provides a critical check point for the systematic exploration and discovery of disordered functional materials, expanding the scope of generative modeling in materials science.
\end{abstract}

\section{Introduction}The discovery of new materials plays a vital role in numerous fields of science and is crucial to develop the next generation of materials. As crystals are the foundation of various materials, crystal structure prediction (CSP) greatly influences future discovery of new materials. Traditionally, the idea of CSP is to return a 3D structure of a compound based on its composition\cite{desiraju2002cryptic}. The ability to accurately and efficiently generate these structures paves the way for the discovery and design of new materials, thereby having considerable impact in many scientific fields\cite{oganov2019structure}. 
Recent advances in CSP within the inorganic crystal domain has encouraged the use of generative models for exploring this vast material space. Various strategies have been utilized, such as Variational auto-encoders (VAE)\cite{CDVAE}, diffusion models\cite{diffcsp, diffcsp_plus,mattergen,cornet2025kinetic}, transformer models\cite{antunes2024crystal,kazeev2024wyckofftransformer}, flow-based crystal generative models\cite{flowmm,flowllm}, and generative adversarial networks (GAN)\cite{kim2020generative}. All of these have shown promising results in the inverse design of inorganic crystalline materials.

Despite the relative success of these models, they all rely on graph-based or invertible feature representations of atoms occupying symmetry-defined sites in the unit cell (the repeating periodic unit) of a crystalline solid. Such representations are unfeasible when describing inorganic crystal structures with site disorder, where (1) either certain sites are occupied randomly by different atomic species, or (2) some atomic sites are unoccupied (vacancies) causing imperfections in the crystal. 
Such disordered inorganic crystals are a crucial consideration as they are very common in functional materials, with ion-mixing or doping strategies being widely applied to enhance functional properties and stability in materials. Examples include solid solutions like the famous Mas/Oregon blue pigment \ce{
YIn_{1-x} Mn_xO3}\cite{smith2009mn3+}, many perovskite-structured materials with cation substitutions\cite{ning2023disorder,chu2023cation}, fast-ion conductors (e.g. \ce{$\alpha$-AgI} where the mobile ions occupy multiple sites within the lattice)\cite{funke2015low}, doped thermoelectric materials like \ce{(PbTe)_{1-x} (PbSe)_x}\cite{wang2013criteria}, as well as electrodes for batteries\cite{zhong2024effect,wang2024disordered}.

To describe such ubiquitous disordered inorganic crystals computationally, it is often necessary to use multiple repetitions of the unit cell, called "supercells". This approach enables the partial occupation of atoms to be distributed among the corresponding symmetrical sites.
However, this method, due to inherent randomness in assigning the partial occupancy to the symmetrical sites of the finite-sized supercell, fails to accurately capture the true nature of the disorder in the crystal structure. To account for issue, traditionally, disordered inorganic crystals are modeled with linear clustering methods, along with Monte-Carlo methods\cite{clease,su2024first}. This approach aims to sample the entire configurational space of the supercell, generating a representative set of supercells with various assignments of the partial occupancy to the symmetrical sites. These sampled configurations approximate the nature of the disordered crystal structure. 

Two primary challenges associated with using this method, especially when aligned with generative models, are: (1) the need to generate hundreds of supercells to describe a single disordered crystal structure, and (2)  the breakdown of symmetry in the crystal structure upon assigning partial occupancy to specific sites, complicating the cell's representation. For smaller cases, it is possible to train a generative model on a sample set of supercells, as demonstrated in\cite{yong2024dismai} and\cite{mattergen}. However, generalizing such models to encompass the diverse range of inorganic disordered crystals is impractical.

A more appropriate approach is to cluster all symmetry-equivalent atomic sites in the crystal structure together, and treat each group of sites independently. This approach enables the explicit modeling of partial occupancies while preserving the site-symmetry associated with each group of crystallographically-equivalent positions, known as Wyckoff sites. In this work, we present Dis-GEN, a framework for representing disordered inorganic crystals based on Wyckoff sites. Our framework enables the training of a VAE model to achieve CSP for disordered crystals. This representation explicitly incorporates partial occupancies of atomic species by encoding both space group symmetries and Wyckoff site symmetries. In our framework, equivariance is defined empirically: any spatial operation—such as rotation or translation—that alters the Wyckoff-based encoding is effectively captured by the symmetry and reflected in the model's reconstruction behavior. Additionally, the model exhibits invariance in predictive tasks, as global spatial operations do not change the physical properties or symmetry information of the crystal structure. In this paper, we start by introducing the representation of disordered crystal structures, followed by the VAE used for generating the structures, and finally the reconstruction and evaluation error.

\section{Crystal representation}
\subsection{Crystal description}
Crystals are highly structured solid materials defined by a repeated arrangement of atoms in space\cite{ai4science2023crystal}. The atomic pattern that periodically repeats itself is called a motif. The parallelepiped containing the motif, which defines each periodicity in the 3D space of the motif, is called a unit cell. A $(u\times v \times w) | u,v,w\in\mathbb{N}$ repetition of the unit cell is called a supercell. Theoretical crystallography has developed methods to systemically describe the endless combinations of crystals using lattice parameters, space groups and Wyckoff sites.

A lattice $\Lambda$ is an infinite set of points defined by the sum of a set of linearly independent primitive lattice vectors, $\Vec{\mathbf{a}}_i \in \mathbb{R}^n$: $ \Lambda = \{R_{[m_i]}^n= \sum^n_i m_i\Vec{\mathbf{a}}_i \}$, where $m_i\in \mathbb{Z}$. In 3D, the lattice can be described by the repeated motif and 6 parameters: $a,b,c$ determines the length of each dimension and $\alpha,\beta,\gamma$ determines the angle between each dimension\cite{simon2013oxford}.

Space groups describe the symmetry operations that the crystal can undergo while preserving the motif within the crystal lattice. This is described as a tuple $(\mathbf{S},\Vec{\mathbf{t}})$, with $\mathbf{S}\in \mathbb{R}^{n\times n}$ being the symmetry matrix and $\Vec{\mathbf{t}} \in \mathbb{R}^n$ being the translation vector. An atomic position maps $\Vec{\mathbf{x}}\in \mathbb{R}^n$ a vector to $\mathbf{S}\Vec{\mathbf{x}}+\Vec{\mathbf{t}}$. In 3D, all space groups are numbered into 230 types, with the first group considered the unsymmetrical group\cite{ai4science2023crystal,Souvignier2016}.

The complete crystal pattern can be reconstructed from the motif and the unit cell, and thus it is sufficient to focus on the atoms inside a restricted space\cite{Souvignier2016}.
\begin{definition}
\label{def:orbit}
    For a space group $\mathcal{G}$ acting on three-dimensional space, $\mathbb{E}^3$ the (infinite) set:
    \begin{equation*}
        \mathcal{O} = \mathcal{G}(X):=\{ g(X)|g\in \mathcal{G} \}
    \end{equation*}
    is called the orbit of $X$ under $\mathcal{G}$. The orbit of point $X$ is the smallest subset of $\mathbb{E}^3$ that contains $X$ and is closed under the action of $\mathcal{G}$.
\end{definition}

Every point in $\mathbb{E}^3$ belongs to exactly one orbit within the space group. The orbit partitions the direct space into disjoint subsets, meaning that an orbit is completely defined by its point in the unit cell, as translating the unit cell by the space group symmetry covers $\mathbb{E}^3$. To account for the case where two symmetry operations map $X$ into the same point, we define subsets within the space group, termed site-symmetry groups $\mathcal{S}\in \mathcal{G}$, which define symmetry operations for $X$ within the space group. The site-symmetry group of a point $X$ is a finite subset of the space group, which is isomorphic to a subset of the point group $\mathcal{P}$ of the space group. The relation between site-symmetry groups of points in the same orbit is the definition of the Wyckoff site\cite{Souvignier2016}:
\begin{definition}
\label{def:site_sym}
    Two points $X$ and $Y$ in $\mathbb{E}^3$ belong to the same Wyckoff site with respect to $\mathcal{G}$ if their site-symmetry groups $\mathcal{S}_X$ and $\mathcal{S}_Y$ are conjugate subsets of $\mathcal{G}$. In particular, the Wyckoff site containing a point $X$ also contains the full orbit $\mathcal{G}(X)$ of $X$ under $\mathcal{G}$.
\end{definition}
Note that it is the definition of Wyckoff sites that any points related by symmetry operations of the space group belong to the same site. The Wyckoff sites themselves are defined using three parameters:
\begin{itemize}
\item Wyckoff letter, which defines the site-symmetry group for a Wyckoff site. It is labeled in alphabetical order, starting with '$a$' for a position with a site-symmetry group of highest site-symmetry. 
\item Wyckoff multiplicity, which defines the number of points in an orbit for a Wyckoff site.
\item Fractional coordinates, which define the real position in the crystal lattice for which symmetry operations can be acted upon.
\end{itemize}

Each Wyckoff site is occupied with a number of atoms equal to the Wyckoff multiplicity. If a Wyckoff site is occupied with one type of atom, it is considered ordered, and if the Wyckoff site is occupied with several types of atoms, it is considered disordered. In this work, an inorganic material is termed disordered if one or more Wyckoff sites are disordered and we do not consider amorphous materials in this work.

\subsection{Representation of disordered inorganic crystals}
\begin{figure}[t!]
    \centering
    \includegraphics[width=1\linewidth]{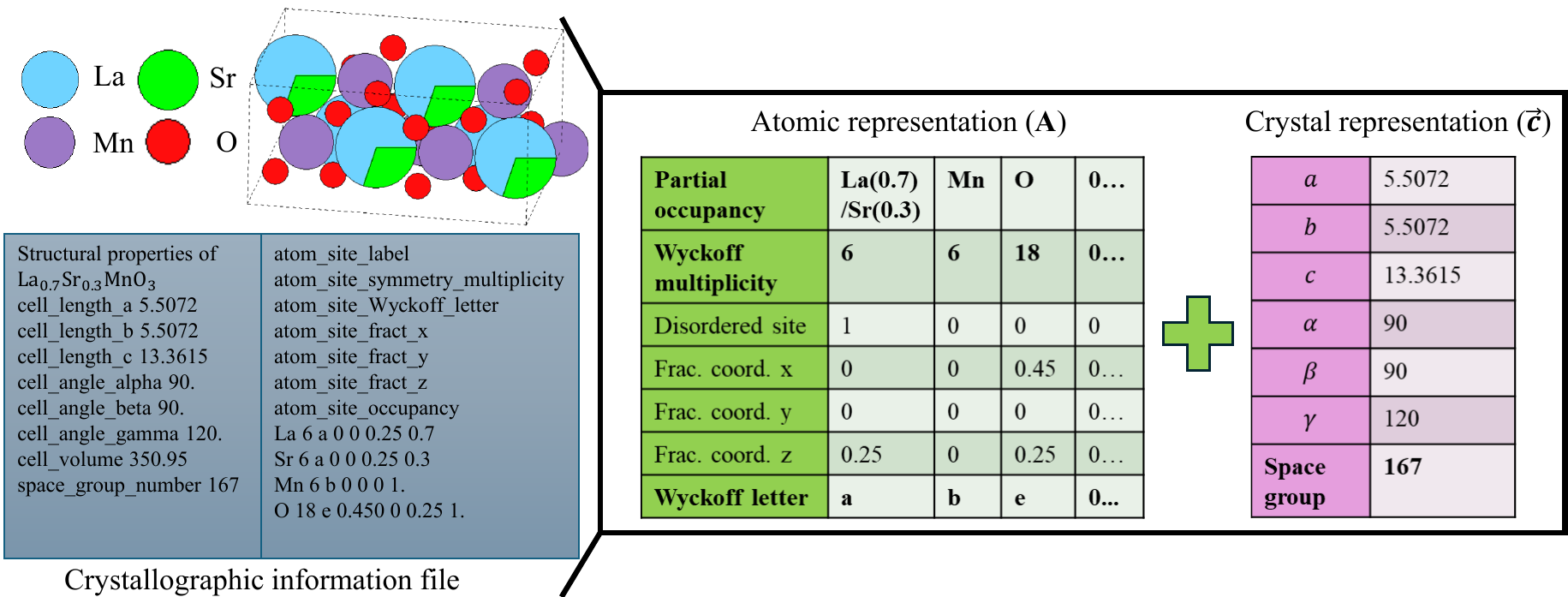}
    \caption{Structural representation of crystals. A Crystallographic information file of a crystal is transformed into a matrix $\mathbf{A}$ describing the atomic representation and a vector $\Vec{\mathbf{c}}$ describing the crystal representation. The bold numbers and letters represents One-hot encoding in the representation. The highlighted colors in the representations denote the labels for each rows.}
    \label{fig:Struc_rep}
\end{figure}

From the crystallography description provided in a Crystallographic Information File (CIF), a disordered inorganic crystal is represented by a matrix $\mathbf{A}$ and a vector $\Vec{\mathbf{c}}$, as illustrated in \cref{fig:Struc_rep}. The atomic configuration is represented by $\mathbf{A}$, while the crystal itself is represented by $\Vec{\mathbf{c}}$. 

Each column in $\mathbf{A}$ represents the different Wyckoff sites in the crystal, while the rows describe different properties within the Wyckoff sites. The first set of rows describes the partial occupancy; the second set describes the Wyckoff multiplicity; the third set describes (1) an indicator for a disordered sites and (2) the fractional coordinates; the last set of rows describes the Wyckoff letter. The partial occupancy, Wyckoff multiplicity and Wyckoff letter are all One-hot encoded with the row number equal to their respective value, while the fractional coordinates are divided into the $x$, $y$ and $z$ components for each row. The disordered site indicator is a binary term; 1 if a Wyckoff site is disordered and 0 if not. Zero padding is added as additional columns to allow matching of crystals with fewer Wyckoff sites to those with the highest number of Wyckoff sites, such that $\mathbf{A}$ is the same size for all crystals. A corresponding zero padding indicator is added to the representation for the partial occupancy, the Wyckoff multiplicity and the Wyckoff letter, such that it can identify whenever a Wyckoff site exists. The first 6 entries in $\Vec{\mathbf{c}}$ consist of the 6 lattice parameters $a,b,c$ and $\alpha,\beta,\gamma$, while the last 230 entries consist of the One-hot encoded space group.

The use of Wyckoff sites and space groups to represent materials for CSP has previously been done for VAE\cite{zhu2024wycryst} and transformer models\cite{antunes2024crystal,kazeev2024wyckofftransformer}. However, to the best of our knowledge, no generalized representation has been developed to account for disordered Wyckoff sites with partial occupancy. Recently, Ref. \citenum{mattergen} demonstrated an approach for disordered structures, but this is limited to a singular type of disorder where two atoms swap positions during the generation of the crystal structure. Their representation does not incorporate partial occupancy of atoms, rendering an incomplete investigation of important ion-mixing and doping practices. By contrast, our representation, combined with the VAE model, provides a novel and previously unexplored framework for the generation of the vast majority of disordered materials, enabling a more comprehensive analysis of disordered crystals.

\section{Disordered VAE}
\subsection{Data}
\begin{figure*}
    \centering
    \includegraphics[width=1\linewidth]{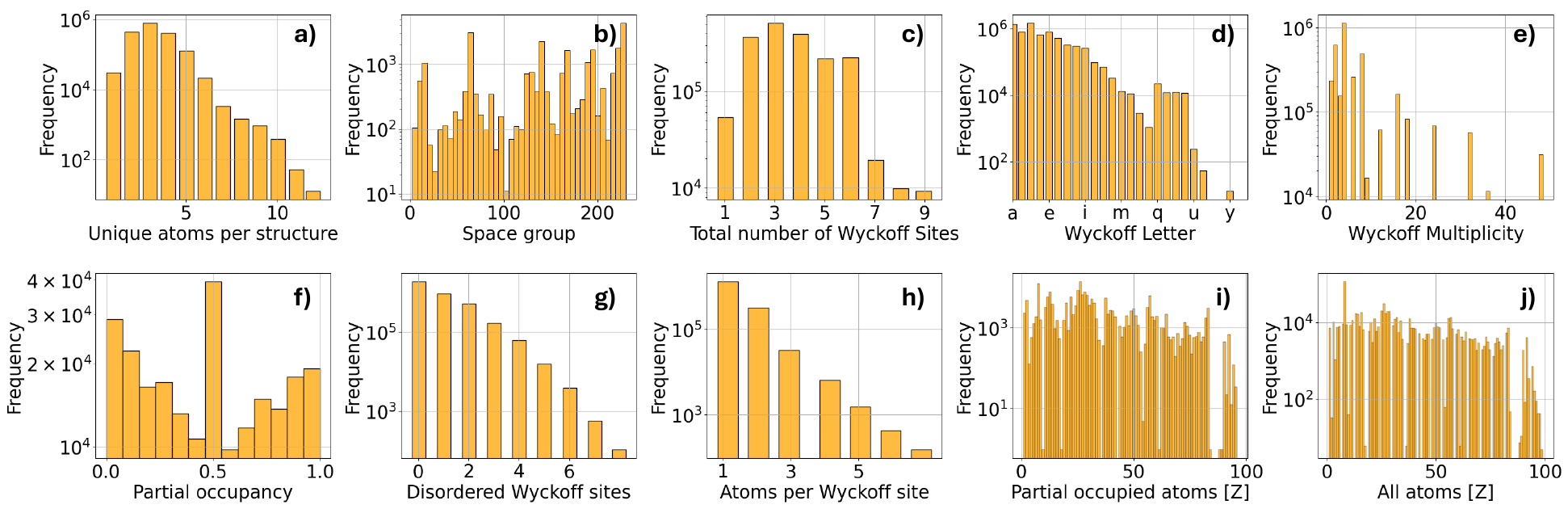}
    \caption{The data distribution after filtering 138,692 structures from the Inorganic Crystal Structure Database (ICSD), with 72,855 structures having partial occupied Wyckoff sites and 65,837 structures having no partial occupied Wyckoff sites. a) Number of unique atoms per structure. b) Space group per structure. c) Number of Wyckoff sites per structure. d) All Wyckoff letters. e) All Wyckoff multiplicities. f) Partial occupancy of disordered Wyckoff sites. g) Number of disordered Wyckoff sites per structure. h) Number of atoms per Wyckoff site. i) The atoms partially occupying the disordered Wyckoff sites. j) All atoms presented in the Wyckoff sites. }
    \label{fig:Data_hist}
\end{figure*}
The Inorganic Crystal Structure Database (ICSD)\cite{icsd} consists of around 229,487 precise experimental inorganic crystal structure entries, all of which are experimental inputs and verified, with 106,970 (45.6\%) of the entries as disordered inorganic crystals, 122,517 (52.2\%) as ordered inorganic crystals, and 4966 (2.2\%) of the entries being infeasible structures not able to be read using Pymatgen\cite{peyret2023electrochemical}. Being an extensive collection of synthesized disordered inorganic crystals, the ICSD represents an ideal dataset for training generative models aimed at disordered Wyckoff sites, even in the absence of computed physical or chemical properties. To date, the disordered crystals presented in ICSD have only been utilized as validation measures for rougher generative models, instead of being actively used as training data\cite{zhu2023predicting, mattergen}. 

To ensure stable training of Dis-GEN, we reduce the complexity of the crystal structures, by considering experimental practice for doing simpler experimental synthesis. We exclude all structures belonging to the first symmetry group (P1) (536 structures), since these are crystals are assigned a unit cell without any symmetrical operations, making it infeasible to consider using a Wyckoff notation. We also exclude structures that contain: (a) rare atoms with a periodic number higher than 100 (7,515), (b) more than nine Wyckoff sites (53,586 structures), (c) greater than 50 in singular Wyckoff multiplicity (2,756 structures), (d) more than six disordered Wyckoff sites (25,671 structures), (e) instances where a Wyckoff site is occupied by more than six distinct atom types (82 structures), and (f) structures with atomic charge state as the partial occupancy (649 structures). We do not exclude duplicated entries from ICSD, as doing so would introduce a different form of bias by favoring one experimental report over another. Although such duplicates may skew representation toward certain structures, retaining them preserves the diversity and integrity of the experimental record.

The data distribution after filtering (138,692 structures) is explored in \cref{fig:Data_hist}, where we see a bias towards ordered Wyckoff sites and fewer total Wyckoff sites - as expected due to the higher number of simpler crystals in the data set. In this subset, 72,855 of the crystal structures contain one or more Wyckoff sites with partial occupancy, while 65,837 of the crystal structures have no Wyckoff sites with partial disorder. The bias toward crystals with fewer total Wyckoff sites justifies our decision to reduce the data set based on the total number of Wyckoff sites. Additionally, the use of extra zero padding would increase the complexity of the representation, which is not worthwhile since it would only accommodate a minor subset of crystals - this argument also holds for most of the other choices.

The ICSD is the most comprehensive and reliable experimental database of inorganic crystal structures to date, with all entries curated and verified by human experts. Leveraging this dataset allows us to generate structures that more closely reflect experimental observations, as our model is trained on experimentally measured unit cell parameters and site occupancies, rather than on values derived from density functional theory (DFT). In contrast, existing generative models\cite{mattergen,CDVAE,diffcsp,diffcsp_plus} typically assume full site occupancy (i.e., site occupancy = 1), which is incorrect for many entries in the Materials Project\cite{MP_database} and rely on DFT-optimized unit cell parameters that often deviate from experimental measurements.

\subsection{Model}
\label{sec:model}
\begin{figure*}
    \centering
    \includegraphics[width=1\linewidth]{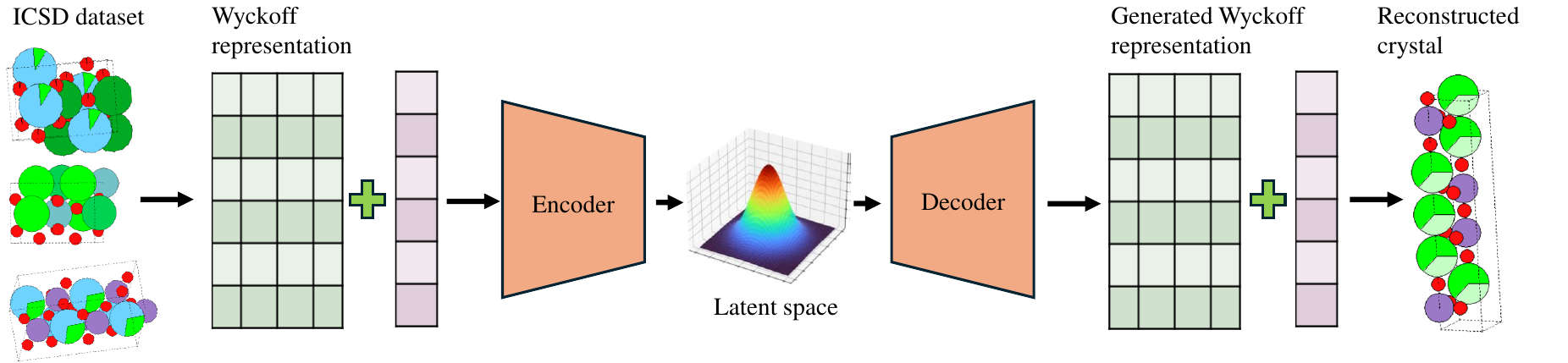}
    \caption{Schematic of the Variational auto-encoder (VAE) used in this work. All the materials gathered from ICSD are represented using the representation from \cref{fig:Struc_rep}. This representation is encoded in batches and a Gaussian-based latent space is created, from which the decoding is based upon. The decoding returns a structural representation, from which the disordered structure is reconstructed. }
    \label{fig:VAE}
\end{figure*}
The main objective of the VAE is to learn the distribution of disordered inorganic crystals from the dataset, to eventually enable crystal structure generation. This procedure is illustrated in \cref{fig:VAE}. Firstly, the VAE needs to encode the representation from \cref{fig:Struc_rep} into the latent space. This is done through a convolution neural network (CNN) with three convolution layers for the atomic representation, and an Multilayer Perceptron (MLP) with two linear layers for the crystal representation. The output of the two networks are combined into $Z_{mean}\in \mathbb{R}^n$ and $Z_{var}\in \mathbb{R}^n$, which parameterize the Multivariate Gaussian distribution in the latent space. Secondly, the decoder is trained to generate samples from the latent space and reconstruct them into the structural representation. This is done through another CNN for the atomic representation and an MLP for the crystal representation. At the end layer of the CNN, the output is divided into the parameters, partial occupation of atoms, Wyckoff multiplicity, disordered site indicator, fractional coordinates and Wyckoff letter, each of which have their own loss function. At the end layer of the MLP for the crystal representation, the output is divided into the lattice parameters and the space group, which similarly have their own loss function.

In total, 7 loss functions are used for the reconstruction of the inorganic crystal, representing the 7 properties constructed during decoding (lattice parameters, space group, disordered site indicator, Wyckoff letter, Wyckoff multiplier, fractional coordinates and partial occupancy at each atomic site). This loss, along with a Kullback-Leiber (KL) divergence loss $\mathcal{L}_{KL}$, constructs the total reconstruction loss. The total reconstruction loss encourages the model to generate the output data as closely as possible to the input data. A detailed description of the loss function is presented in \cref{app:Training}

To illustrate the capability of Dis-GEN, the filtered ICSD, as represented in \cref{fig:Struc_rep}, is used to train a VAE model. The dataset (138,692 structures) is randomly split into a validation set (10\%) and test set (20\%). The specifics of the VAE model are shown in \cref{app:Training}.

\section{Generation of disordered inorganic crystals}
\subsection{Reconstruction error}
\begin{table*}[htbp]
    \centering
    \caption{Reconstruction error of the test set for the lattice parameters (MAE), space group (Accuracy) and disordered site indicator (Accuracy).}
    
    \resizebox{0.4\linewidth}{!}{%
    \begin{tabular}{lc}
    \toprule
    \bf Parameters &\bf  Dis-GEN \\
    \midrule
    ($a$,$b$,$c$) [Å]  & (0.06, 0.06, 0.10) \\
    ($\alpha$,$\beta$,$\gamma$) [\textdegree]    & (0.02, 0.05, 0.28) \\
    Space group & 99.4\%\\
    Disordered site & 99.8\% \\
    \bottomrule
    \label{table:result_lattice}
    \end{tabular}
    }
    
\end{table*}

From the test set, we assess the reconstruction errors associated with the encoding and decoding of Dis-GEN.
As shown in \cref{table:result_lattice}, the reconstruction error of the lattice parameters, in mean absolute error (MAE), is small - as are the errors related to the space group and the disordered site indicator. Notably, the reconstruction losses for the lattice parameters $c$ and $\gamma$ are higher than those for the other parameters, which is consistent with their greater variability across ICSD.

Regarding the Wyckoff site parameters presented in \cref{table:result_wyckoff}, we achieve high accuracy for both the Wyckoff letter and Wyckoff multiplicity, for both the ordered and disordered Wyckoff sites. Similarly, the reconstruction error of the fractional coordinates remains small. For the partial occupancy, we use two different error metrics for the ordered and disordered cases. For the ordered structures, we can report an accuracy score - a single atomic species occupies any Wyckoff site, making it a classification problem (correctly or wrongly occupied). However, for the disordered structures, several atom species can occupy the same Wyckoff sites, making it a regression problem and hence why we report a corresponding MAE. We achieve a satisfactory result considering that the majority of inorganic structures exhibit occupancies greater than 5\%, as illustrated in \cref{fig:Data_hist}. In future training of the model, a higher accuracy would be a valuable improvement, as experimental studies have demonstrated that even incorporating low concentrations of elements can enhance the functional properties of materials\cite{ahaliabadeh2022extensive,chen2019effects}. Remarkably, the model does not overfit to either disordered or ordered Wyckoff sites, despite the bias present in the dataset as shown in \cref{fig:Data_hist}. This is a significant achievement, as it demonstrates that the model is not biased towards any specific group of Wyckoff sites. Instead, it maintains a balanced representation, allowing for accurate predictions across the entire range of Wyckoff site configurations.

\begin{table*}
\caption{Reconstruction errors of the test set for the parameters directly related to the disordered and ordered Wyckoff sites. The partial occupancy is presented with a MAE for the disordered sites, and an accuracy for the ordered sites. Accuracy is also used for the Wyckoff letter and Wyckoff multiplicity, while MAE is used for the fractional coordinate.}
\label{table:result_wyckoff}
\centering
\resizebox{0.85\linewidth}{!}{%
\begin{tabular}{lcc}
\toprule
\bf Parameters  &\bf  Disordered Wyckoff sites &\bf  Ordered Wyckoff sites \\
\midrule
Partial occupancy & 0.06 (MAE) & 98.6\%  \\
Wyckoff multiplicity & 99.6\% & 99.8\% \\
Wyckoff letter & 99.5\% & 99.7\%  \\
Frac. coordinate [Å] & 0.07 & 0.07 \\
\bottomrule
\end{tabular}
}
\end{table*}

The Wyckoff site is determined by the site-symmetry and space group. Consequently, for a given Wyckoff letter and space group, the corresponding Wyckoff multiplicity is uniquely defined. The Wyckoff letters are also uniquely determined by the space group. Leveraging this, we evaluate the representations throughout the VAE by comparing the predicted Wyckoff multiplicity with the reference values determined by the space group and Wyckoff letter, along with comparing the predicted Wyckoff letter with the reference letter determined by the sapce group. This evaluation defines the symmetry-matching accuracy (SMA), and it was found that 98.4\% of the reconstructed test set predicted Wyckoff multiplicity and Wyckoff letter following the reference values. This SMA remains similar to the reconstructed accuracy of the Wyckoff multiplicity value in \cref{table:result_wyckoff}, implying the symmetry is preserved throughout the VAE. 

A visualized representation of the reconstruction error can be considered in \cref{app:Recon_error}.

\subsection{Crystal structure generation}
\begin{figure}[b]
    \centering
    \includegraphics[width=0.5\linewidth]{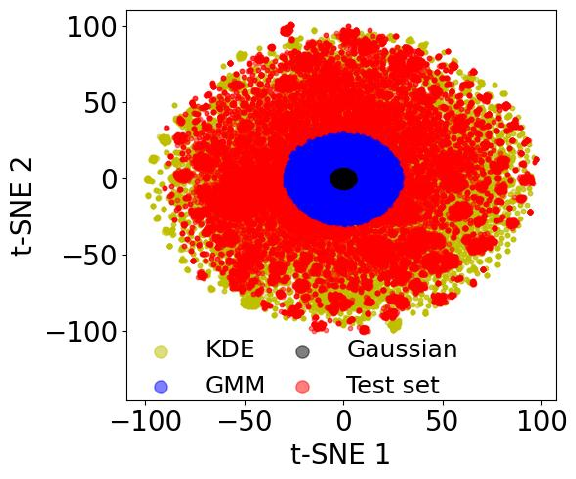}
    \caption{The latent space of the test set compared to a Multivariate Gaussian sampling, along with a KDE\cite{kde} sampling and a GMM\cite{gmm} sampling, which are both trained on the latent space of the training set. The sample set for all three models are of the same size as the test set, consisting of 27,008 structures.}
    \label{fig:latent_space}
\end{figure}
Crystal structure generation inherently carries the risk of generating physically meaningless structures. However, our representation is grounded in both crystal and site-symmetry, which allows us to minimize this issue during the reconstruction process. Specifically, by leveraging symmetry constraints, we can effectively avoid the formation of unreasonable inorganic crystal structures. For this reason, the SMA is determined during the reconstruction process and non-symmetry-validated structures are removed. Unreasonable structures containing close to zero lattice constants or zero Wyckoff multiplicities are also removed during reconstruction. The reconstruction process begins by identifying the crystal lattice and space group. The empty Wyckoff sites are then identified using the reconstructed Wyckoff letter - from which the partial occupancy, fractional coordinates and Wyckoff sites can additionally be determined for the non-empty Wyckoff sites. From this, we have all the information needed to generate the symmetrized structure. 

Even though the reconstruction errors of the fractional coordinates are found to be minor in \cref{table:result_wyckoff}, it can still cause an overlap of atoms when the symmetry operations, given by the Wyckoff sites and space groups, are considered. To avoid this error, we use PyXtal\cite{pyxtal} to refine the fractional coordinates, to align to the nearest fractional coordinate given by the Wyckoff sites and space group. This approach is performed similarly to the procedure used by SymmCD\cite{levy2025symmcd}.

Because some Wyckoff sites exhibits a large Wyckoff multiplicity (as shown in \cref{fig:Data_hist}), some of the generated structures becomes quite large. This increased structural size makes the direct use of PyXtal\cite{pyxtal} impractical for post-processing. Instead, we generate the symmetrized structures directly in CIF format, enabling efficient and scalable structure generation.

To perform crystal structure generation from the latent space of Dis-GEN, it is vital to accurately characterize its posterior distribution. The KL divergence loss encourages the formation of a smooth Multivariate Gaussian distribution; however, this comes at the cost of an increased reconstruction loss. Given the complexity of representing disordered crystals, achieving a perfectly Gaussian latent space is inherently challenging. To further analyze the latent space, we use t-SNE\cite{tsne} to reduce the dimensionality of the latent space to two dimensions, facilitating visual comparisons of the test-set latent space. 

To sample beyond the train and test sets, it is crucial to estimate parts of the latent space using different estimators. Beyond the conventional Multivariate Gaussian sample, the Kernel Density Estimation (KDE)\cite{kde} and the Gaussian Mixture Model (GMM)\cite{gmm} are trained on the training dataset to estimate the overall latent space. These are visualized along with the latent space of the test set in \cref{fig:latent_space}. Visually, the KDE model offers the most accurate approximation of the latent space, providing a reasonable representation of its diverse distribution, while the GMM\cite{gmm} and the Multivariate Gaussian distribution show less satisfactory results. This observation is further supported in \cref{table:recon}, where the generation error (defined as the percentage of sampled structures discarded during the reconstruction process) and SMA highlights the differences between the three models.

\subsection{Evaluating generated structures}

Evaluating disordered structures is inherently challenging due to their partial occupancies, which makes direct structural comparisons infeasible. Comparing structures with differing partial occupancies is effectively equivalent to comparing different compositions within the same crystal family, rendering traditional structural similarity metrics imprecise. Furthermore, the presence of partial occupancies significantly complicates the computation of functional properties, such as the formation energy and/or the energy above the hull, as it necessitates the sampling and averaging of over hundreds of possible configurations. As a result, commonly used metrics such as the stable-unique-novel (S.U.N) metric\cite{CDVAE,kazeev2024wyckofftransformer,mattergen} are not directly applicable to disordered inorganic materials.

\begin{table*}[t!]
\caption{ Evaluation of the three latent space estimators using four metrics: generation error, SMA, structural validity metric (validity), and the charge neutrality filter. All estimators are evaluated using a dataset of the same size as the test set, consisting of 27,008 structures. The reported values for the generation error and SMA correspond to the accuracy, defined as the fraction of structures that passed each respective filter over all generated structures. In contrast the validity, and charge neutrality is computed over all structures not removed during the reconstruction process. }
\label{table:recon}
\centering
\resizebox{0.9\linewidth}{!}{%
\begin{tabular}{lcccc}
\toprule
\bf Latent space  &\bf  Generation error &\bf  SMA &\bf  Validity&\bf Charge neutrality \\
\midrule
KDE & 99.94\% & 98.79\% &96.44\% & 55.28\%\\
GMM & 98.17\% & 20.44\% & 94.03\% & 50.67\% \\
Gaussian & 97.32\% & 8.42\% & 95.03\% & 55.61\%\\
Test set &99.99\% & 98.39\% & 97.53\% & 55.06\% \\
\bottomrule
\end{tabular}
}
\end{table*}

While the S.U.N metric is not applicable to disordered structures, the structural validity criterion metric by Ref. \citenum{CDVAE}—which requires that no two atoms be closer than 0.5Å—can still be employed to evaluate disordered materials. Similarly, a charge neutrality filter proposed by Ref. \citenum{smact} can be applied to ensure chemical plausibility. However, enforcing charge neutrality is more complex in the context of disordered structures, as some entries in the ICSD contain atomic species with multiple charge states, adding an additional source of error as the approach only considers the most likely charge state. Thus, the generation error, SMA, structural validity criterion, and charge neutrality filter are used as appropriate evaluation metrics to screen the generated structures. These filters ensure that only chemically reasonable and physically plausible structures are retained for further analysis.

The evaluation metrics are summarized in \cref{table:recon} for both the test set and the three latent space estimators. While the generation error remains relatively consistent across all estimators and the test set, the SMA is notably highest for KDE, even slightly exceeding the value observed for the test set. This result suggests that sampling from the KDE-estimated latent space promotes empirical symmetry equivariance in the generated structures. The validity is similar for all estimators, indicating a consistent generation when the structures are parsed through the SMA filter. As expected, the charge neutrality filter remains consistently low across all estimators due to the presence of atomic species with multiple oxidation states, which complicates formal charge balancing in disordered structures.

In general, by using KDE to estimate the latent space, a stable sampling methodology is ensured, making exploration of the latent space more robust.

\subsection{Unconditional structure generation}

\begin{figure*}
    \centering
    \includegraphics[width=1\linewidth]{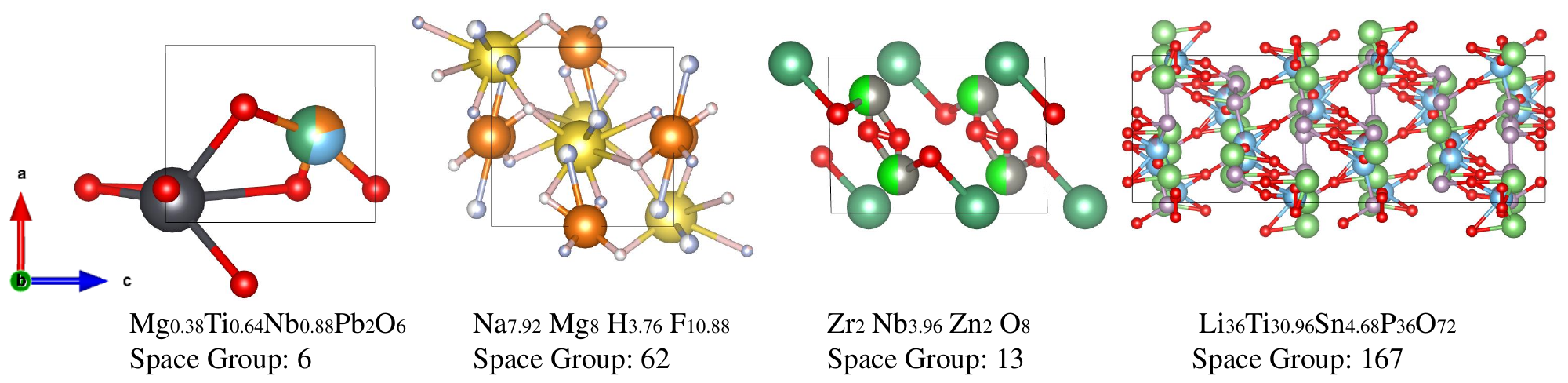}
    \caption{Four generated disordered inorganic crystal structures, using Dis-GEN. All crystals are viewed along the b-axis of the crystal.}
    \label{fig:recon}
\end{figure*}
Using the KDE estimation of the latent space, we can generate disordered structures, with four representative generations illustrated in \cref{fig:recon}. Given the stochastic nature of our sampling process, the generated structures appear reasonable.  The leftmost crystal, \ce{Mg_{0.38}Ti_{0.64}Nb_{0.88}Pb2O6}, suggests a doping strategy involving \ce{Mg}, \ce{Ti}, and \ce{Nb}, elements commonly used to enhance functional materials for catalysis and biomedical applications\cite{cui2014mg,li2021characteristics}. The generated \ce{Na_{7.92}Mg8H_{3.76}F_{10.88}} and \ce{Li36Ti_{30.96}Sn_{4.68}P36O72} structures resemble electrode materials for batteries. The latter in particular closely aligns with known lithium-ion battery anode or cathode materials, featuring doping at the metallic (\ce{Ti}) site, although \ce{Sn} may not be an appropriate dopant\cite{weng2017high}. 
Additionally, the generated \ce{Zr2Nb_{3.96}Zn2O8} structure is structurally related to the \ce{ZrNbO4} alloy\cite{peyret2023electrochemical}, with \ce{Zn} partially occupying the \ce{Zr} site, suggesting a potential doping strategy for this material. 

Despite the promising nature of the generated structures, certain compositions, such as \ce{Na_{7.92}Mg8H_{3.76}F_{10.88}}, appear chemically unreasonable. This indicates the necessity of incorporating chemically intuitive filtering into the generation process to eliminate unrealistic structures. Currently, no established chemical intuition filters exist for disordered crystals. However, we are actively developing such methods, which will assess the likelihood of partial occupations of atoms at symmetrical sites based on factors such as atomic charge and chemical bonding characteristics.

A limitation of using a VAE model for structure generation and CSP is that the latent space is constrained by the distribution of the training set. While KDE is highly effective at interpolating within known distributions, it struggles with extrapolation, which could be addressed by using diffusion models or transformers. This explains why the generated structures in \cref{fig:recon} closely resemble battery electrodes and alloys, since they are among the most investigated materials, particularly in the context of disordered inorganic crystals. 

Despite this, the ICSD dataset encompasses a vast and diverse range of experimentally verified inorganic crystals. This extensive coverage suggests that numerous potential crystal structures remain unexplored in the ICSD distribution, presenting significant opportunities for the discovery of novel functional materials. This is particularly true for disordered crystals, where possible doping configurations are virtually limitless. However, blindly searching through this vast chemical space is inefficient. To guide the search towards promising candidates, a property-oriented approach incorporating physical and chemical constraints is necessary.

However, computing physical or chemical properties of disordered materials presents significant challenges. Accurate modeling of partial occupancies typically requires the construction of large supercells, which are necessary to approximate the configurational ensemble. However, assigning partial occupancy to symmetry-equivalent sites inherently breaks the crystal symmetry, making such calculations computationally expensive and more complex. Furthermore, to generate a property distribution that is comparable to experimental measurements, one must sample a sufficient number of configurations at sufficiently large supercell sizes. Even then, the resulting comparisons may remain inaccurate, as illustrated in \cref{app:Property}. 

Such configuration-dependent property distributions are rarely captured in theoretical materials databases such as the Materials Project(MP)\cite{MP_database}, which limits the applicability of universal machine learning interatomic potentials\cite{batatia2023foundationmace, deng2023chgnet, m3gnet} for disordered systems. Instead, external methods are required to accurately map the configurational space and derive meaningful averaged properties\cite{xie2024machine}. For these reasons, property-oriented searches remain largely unexplored for disordered inorganic crystal structures.

\subsection{Conditional structure generation}
\begin{figure*}[b]
    \centering
    \includegraphics[width=0.9\linewidth]{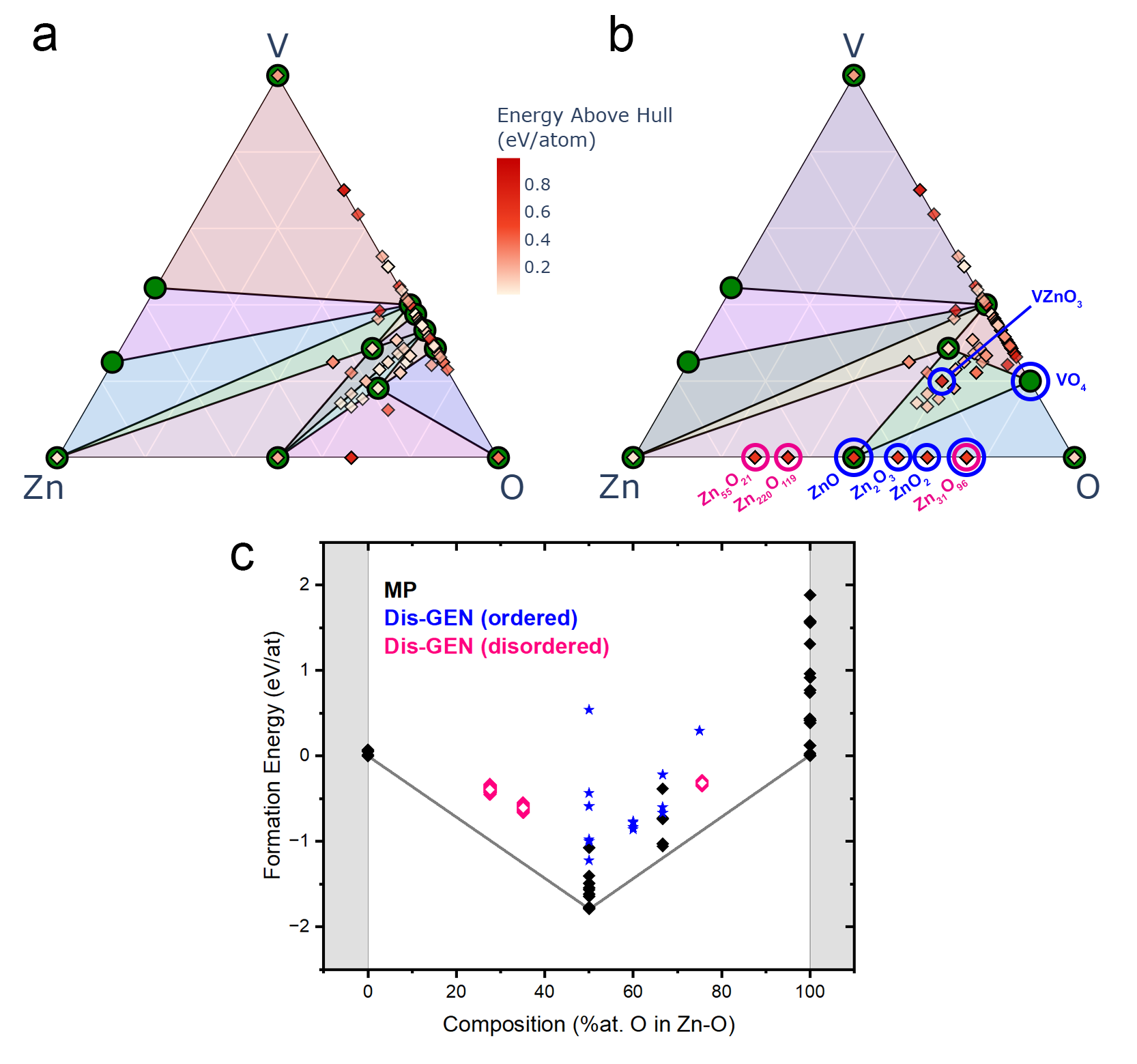}
    \caption{Phase space of the \ce{Zn}–\ce{V}–\ce{O} chemical system, combining structures from the MP\cite{MP_database} and those generated by Dis-GEN. a) Phase space constructed using only MP structures. b) Phase space including both MP structures and highlighted Dis-GEN generated structures. c) Convex hull for the \ce{Zn}–\ce{O} binary system. }
    \label{fig:convex}
\end{figure*}
While it is not possible to condition our generation based on desired properties yet, it is possible to condition based on composition. In our approach, the latent space of the VAE model allows tuning our search to specific compositions, space groups, or even to target certain partial occupations of atoms. By defining specific terms for the desired configurations, we employ gradient descent to optimize the latent sampling $z$ with respect to the target configuration, as illustrated in \cref{app:Condition}. This is feasible because Dis-GEN reconstructs multiple parameters related to both crystal symmetry and atomic representation.

Although it is not feasible to compute physical or chemical properties for a large number of disordered materials, such evaluations are possible for specific compositions of interest. Here, we consider the chemical system \ce{Zn}-\ce{V}-\ce{O} as an example, a recently trending oxide of promise for battery application \cite{Dhn2023} that has been found to favor disordered phases over ordered ones\cite{Haiwen_paper}. While the \ce{ABO3} chemical space has been extensively studied, the \ce{Zn}-\ce{V}-\ce{O} system is underrepresented in the ICSD, making it an ideal case study for testing Dis-GEN and its ability to propose plausible dopants and discover stable configurations within the phase space.

Using Dis-GEN, we generated several promising candidate structures for the \ce{Zn}-\ce{V}-\ce{O} system, by conditioning our search in latent space to these specific atomic species. The  candidates we found are structurally optimized to a maximum force of $0.1$eV/Å using a multi-atomic cluster expansion (MACE) model\cite{batatia2022mace} pretrained on the Open Materials 2024 (OMat24) dataset\cite{barroso2024open}. For ordered phases, a single geometry optimization was sufficient to assess stability. However, for disordered structures, we constructed large supercells and sampled 400 random atomic configurations per candidate in order to reliably estimate the property distribution. 

\cref{fig:convex} illustrates the phase space of the \ce{Zn}-\ce{V}-\ce{O} system, constructed using structures from MP dataset and those generated by Dis-GEN. In comparing \cref{fig:convex}a) and \cref{fig:convex}b), it is clear that Dis-GEN predominantly samples compositions rich in \ce{Zn}-\ce{O} - compositions which are underrepresented in the MP database. Notably, the MP dataset contains no disordered phases, whereas Dis-GEN successfully generates several such candidates. As shown in \cref{fig:convex}c), the generated disordered phases generally lie above the convex hull, indicating thermodynamic instability. 

Both the ordered and disordered phases generated by Dis-GEN are found to be thermodynamically unstable. This may be attributed to the absence of an explicit property-based conditioning in the Dis-GEN model. This example highlights two key challenges in modeling disordered systems: 1) the need for more comprehensive property data for disordered phases, and 2) the development of efficient machine learning interatomic potentials capable of accurately describing such systems. Addressing these challenges will significantly enhance the robustness and predictive power of Dis-GEN, enabling more effective and accurate exploration of the vast and underexplored space of disordered inorganic materials.

\section{Conclusion}

In this work we introduce \textbf{Dis-GEN: Disordered crystal structure generation}, the world's first generative model designed for disordered inorganic crystal structures, leveraging a novel representation for disordered crystals to incorporate partial occupancy at symmetrical sites. To demonstrate the capabilities of Dis-GEN, we utilize the Inorganic Crystal Structure Database (ICSD) - the largest collection of experimentally inorganic crystals - to train a generative model, enabling crystal structure prediction for disordered crystals. Due to the empirical equivariant representation of crystal structures, Dis-GEN inherently filters out symmetry-violating structures, ensuring a symmetry-consistent approach to generating valid structures. 
Moreover, Dis-GEN offers a promising alternative framework to the practically-infeasible property-oriented search, instead targeting compositional similarity for latent space exploration of disordered inorganic structures. 
With Dis-GEN, we initiate the systematic exploration of disordered inorganic crystals, aiming to discover novel structures with potential applications across various scientific fields.


\textbf{Acknowledgements}
A.B., J.M.G.L., and M.H.P. acknowledge support from the Det Frie Forskningsråd under Project ``Data-driven quest for TWh scalable \ce{Na}-ion battery (TeraBatt)'' (Ref. Number 2035-00232B) as well as support from the Novo Nordisk Foundation Data Science Research Infrastructure 2022 Grant: A high-performance computing infrastructure for data-driven research on sustainable energy materials, Grant no. NNF22OC0078009. A.B. also acknowlegde  ADANA project (Grant No. 3164-00297B). S.A. acknowledges support by the UCL-A*STAR Collaborative Programme via the Centre for Doctoral Training in Molecular Modelling and Materials Science (M3S CDT) at UCL. K.H. acknowledges funding from the MAT-GDT Program at A*STAR via the AME Programmatic Fund by the Agency for Science, Technology and Research under Grant No. M24N4b0034. K.H. and A.P.C acknowledges funding from Ministry of Education Tier 1 Award Number: RG138/23.

\textbf{Data and code accessibility}
The data and code based used for this work, will be public upon release. Early access can be permitted by writing to the authors. 

\bibliography{refs}
\newpage

\begin{appendices}

\section{VAE Model specifics}
\label{app:Training}

\begin{figure}[b!]
    \centering
    \includegraphics[width=1\linewidth]{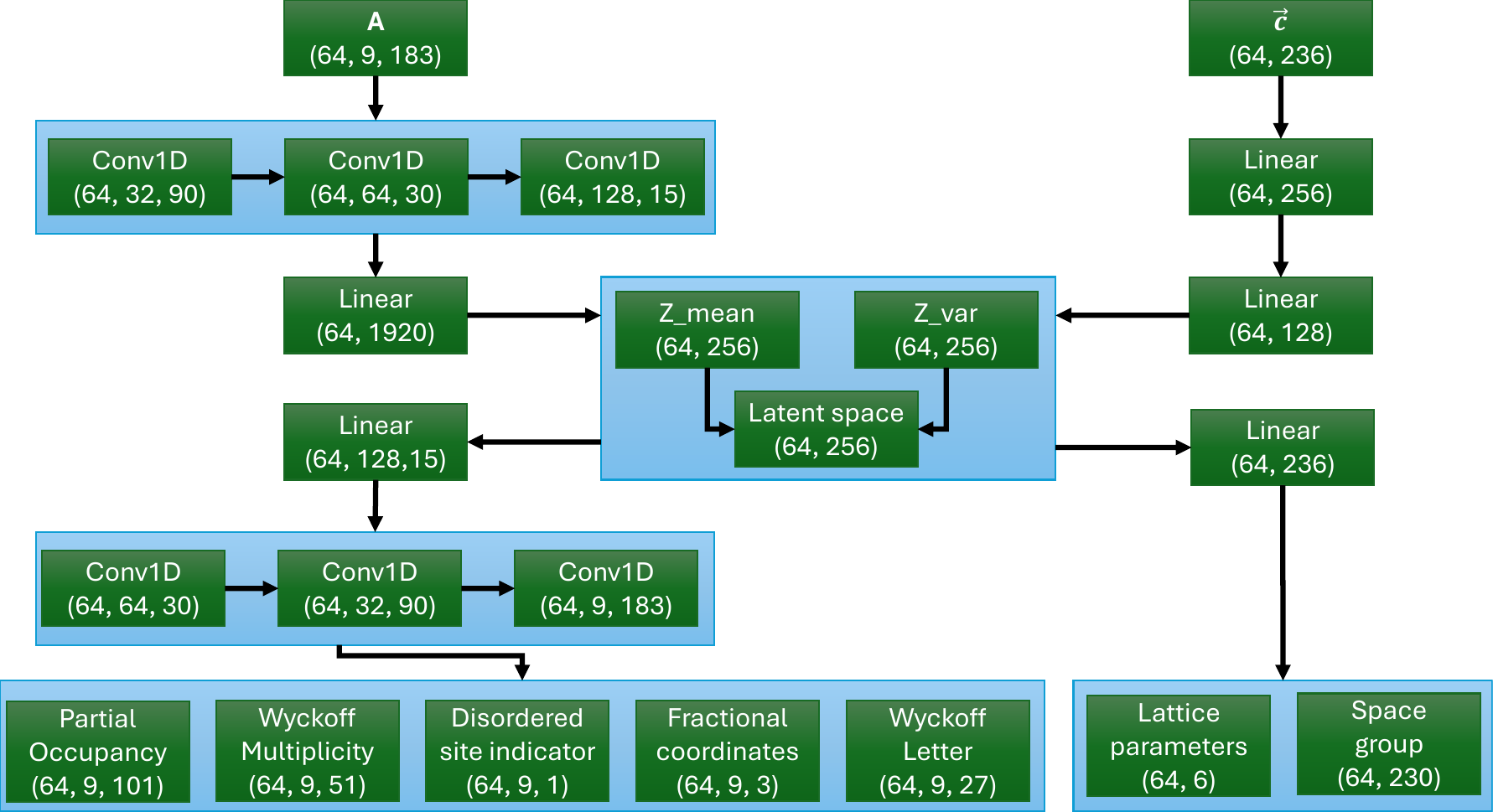}
    \caption{The architecture of the VAE model used for Dis-GEN with a batch size of 64. In the schematic representation, green blocks denote individual layers within the VAE, while blue blocks represent groups of layers. The model accepts two inputs: (1) matrix $\mathbf{A}$, which encodes the atomic representation and (2) a vector $\Vec{\mathbf{c}}$ , which encodes the crystal representation. These inputs are processed separately by the encoder, then concatenated in the latent space and then separated in the decoder. The atomic representation produces five outputs, while the crystal representation yields two two outputs. The ReLU activation function is utilized for the CNN, while the sigmoid activation function is utilized for the disordered site indicator and the softmax activation function is utilized for the partial occupancy, Wyckoff multiplicity, Wyckoff letter and space group. }
    \label{fig:vae_rep}
\end{figure}
The total reconstruction loss function consists of 7 loss functions $\mathcal{L}_{recon}$, along with a Kullback-Leiber (KL) divergence loss $\mathcal{L}_{KL}$.

For the reconstruction loss, two different loss functions are used: 

The cross entropy loss function, which calculates the likelihood of getting the reconstructed distribution $\hat{Q}$, given the input distribution $Q$:
\begin{equation}
    \mathcal{H}(\hat{Q}|Q) = - \sum_i Q(i )\log(\hat{Q}(i))  
\end{equation}
The mean square error (MSE) loss function compares the difference between the input values $Y$ and the reconstructed values $\hat{Y}$:
\begin{equation}
    MSE(Y,\hat{Y}) = \frac{1}{n} \sum_{i=1}^n(Y_i-\hat{Y})^2
\end{equation}
For the parameters in the atomic representation, the loss functions are taken per Wyckoff site and the total loss is a sum of all individual losses. The Wyckoff letter loss $\mathcal{L}_{letter}$, the Wyckoff multiplicity loss $\mathcal{L}_{mult}$ and disordered site indicator loss $\mathcal{L}_{disorder}$ use the cross entropy loss function, while the fractional coordinates loss $\mathcal{L}_{coord}$ and the partial occupancy loss $\mathcal{L}_{occ}$ use the MSE loss function. Note that for partial occupancy, the loss is defined by the difference between the distributions of atoms at a Wyckoff site.

For the parameters in the crystal representation, the loss functions are used directly on the representation. The lattice parameter loss $\mathcal{L}_{lattice}$ uses the MAE loss function, while the space group loss $\mathcal{L}_{spg}$ uses the cross entropy loss function.

The KL divergence loss $\mathcal{L}_{KL}$ is used to shape the latent space into a Multivariate Gaussian distribution. The KL divergence loss calculates the difference between $q(z|X)$, the learned distribution of latent points $z$ given input data $X$, and $p(z)$, the desired Gaussian distribution for the latent points,\cite{vae,kullback1951information}:
\begin{equation}
    \mathcal{L}_{KL} = KL(q(z|X)||p(z))
\end{equation}

During the optimization step of the VAE model, the total loss function
\begin{equation} \label{eq:tot_recon}
\begin{split}
    \mathcal{L} = & \lambda_{KL}\mathcal{L}_{KL}+\lambda_{spg}\mathcal{L}_{spg}+\lambda_{lattice}\mathcal{L}_{lattice} +\lambda_{occ}\mathcal{L}_{occ} 
\\&+\lambda_{mult}\mathcal{L}_{mult}+\lambda_{letter}\mathcal{L}_{letter}
+\lambda_{disorder}\mathcal{L}_{disorder}+\lambda_{coord}\mathcal{L}_{coord}
\end{split}
\end{equation}
is optimized, with $\lambda_i$ as the coefficient for each loss contribution.

\begin{table*}[b]
\caption{Loss coefficients used for training the VAE model used in this paper.}
\centering
\label{table:coeff}
\resizebox{0.4\linewidth}{!}{%
\begin{tabular}{cc}
\toprule
\bf Coefficients  &\bf  Optimal value \\
\midrule
$\lambda_{KL}$ & 1.0 \\
$\lambda_{spg}$ &  10.0 \\
$\lambda_{lattice}$ & 3.0 \\
$\lambda_{occ}$ &  2000 \\
$\lambda_{mult}$ &  1.0 \\
$\lambda_{letter}$ & 1.0 \\
$\lambda_{disorder}$ & 0.1  \\
$\lambda_{coord}$ &  1.0\\
\bottomrule
\end{tabular}
}
\end{table*}

\cref{fig:vae_rep} illustrates the architecture of the VAE model used for generating disordered inorganic crystals. The VAE takes two inputs: (1) a matrix $\mathbf{A}$ describing the atomic representation and (2) a vector $\Vec{\mathbf{c}}$ describing the crystal representation. The model produces seven outputs: partial occupancy, Wyckoff multiplicity, disordered site indicator, fractional coordinates, Wyckoff letter, lattice parameters and space group. To optimize the model architecture, hyperparameter tuning was conducted for the CNN to determine the optimal number of layers, channel dimensions and signal dimensions. Similarly, for the MLP, hyperparameter tuning was performed to optimize the number of hidden layers and nodes. 

\begin{figure}[t!]
    \centering
    \includegraphics[width=0.8\linewidth]{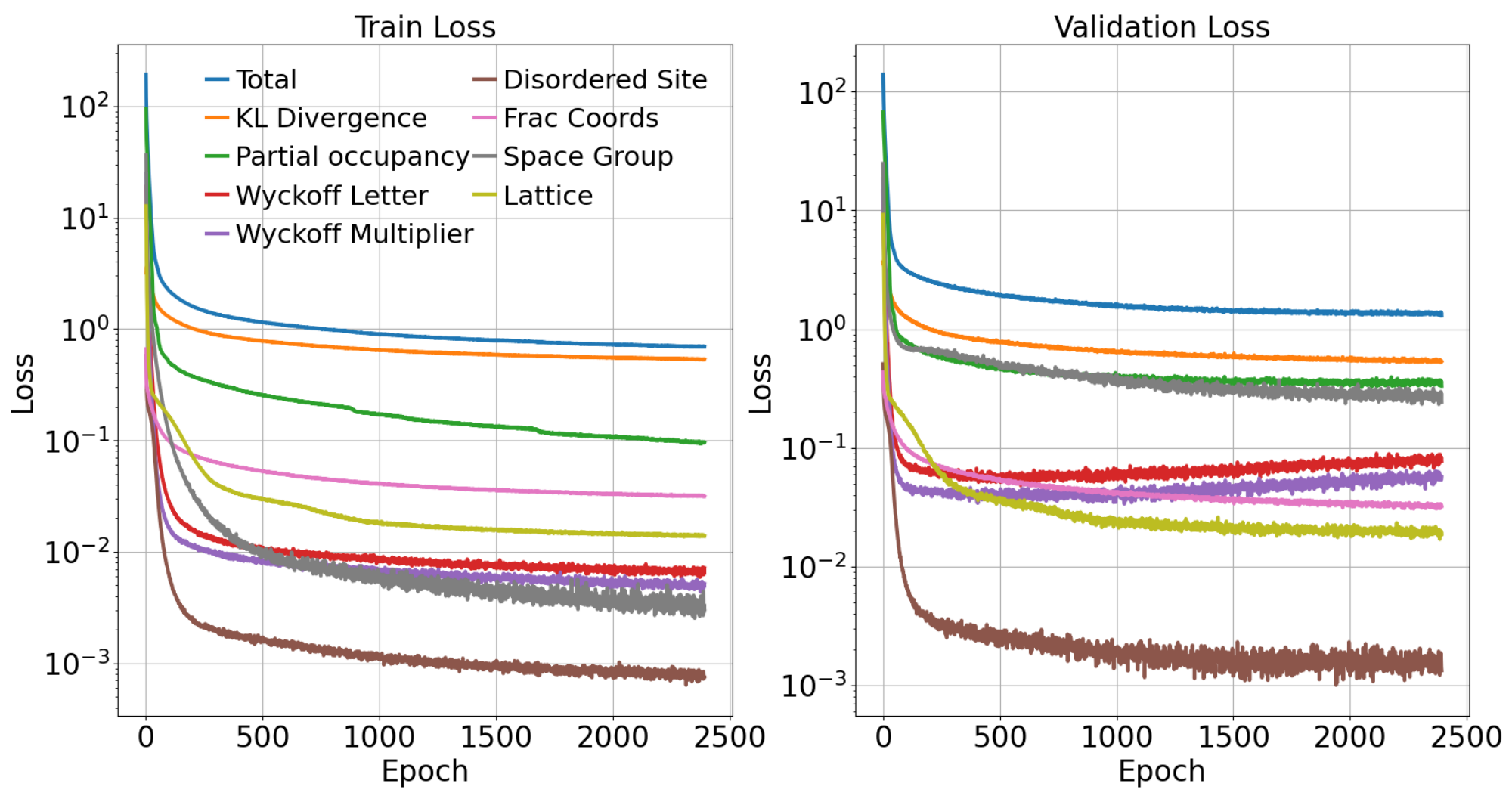}
    \caption{The VAE training curves, with curves color-coded uniformly across the plots.}
    \label{fig:train}
\end{figure}

For the training, the coefficients in the total loss function  (\cref{eq:tot_recon}) were optimized to ensure robust performance on the test set while preventing overfitting, as assessed by validation loss. The final coefficients, presented in \cref{table:coeff}, were determined to provide the most effective balance. Notably, partial occupancy was the most challenging parameter to optimize, resulting in a higher loss coefficient to improve accuracy. In contrast, the difference between the other coefficients are relatively small. The disordered site was the easiest parameter to predict, which justifies its assignment as the lowest coefficient.

Training was conducted using the Adam optimizer\cite{kingma2014adam} with a learning rate of $5\times10^{-6}$ and a batch size of 64. The model was trained for 2500 epochs, with the final VAE model selected based on the lowest validation loss, which occurred at epoch 2349. 

\cref{fig:train} visualizes the training process, where it is evident that the KL loss dominates among the eight loss functions. This behavior is anticipated, as the KL loss must balance the combined effect of all reconstruction-related loss terms. Some variations are observed in specific loss functions that could potentially be avoided by using a lower learning rate. However, these variations are minimal and can be considered negligible.

\section{Visualization of the reconstruction error }
\label{app:Recon_error}
\begin{figure}[h!]
    \centering
    \includegraphics[width=0.45\linewidth]{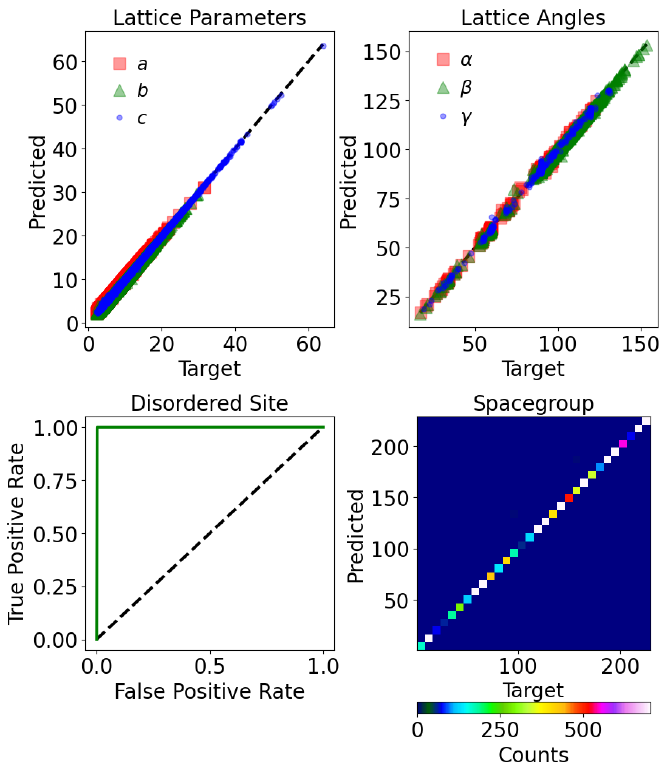}
    \caption{The lattice parameter prediction of the test set along with the disordered site prediction. Note the perfect ROC curve stems from the fact that only 1\% of the Wyckoff sites were misclassified.}
    \label{fig:lattice}
\end{figure}
\begin{figure}[h!]
    \centering
    \includegraphics[width=0.85\linewidth]{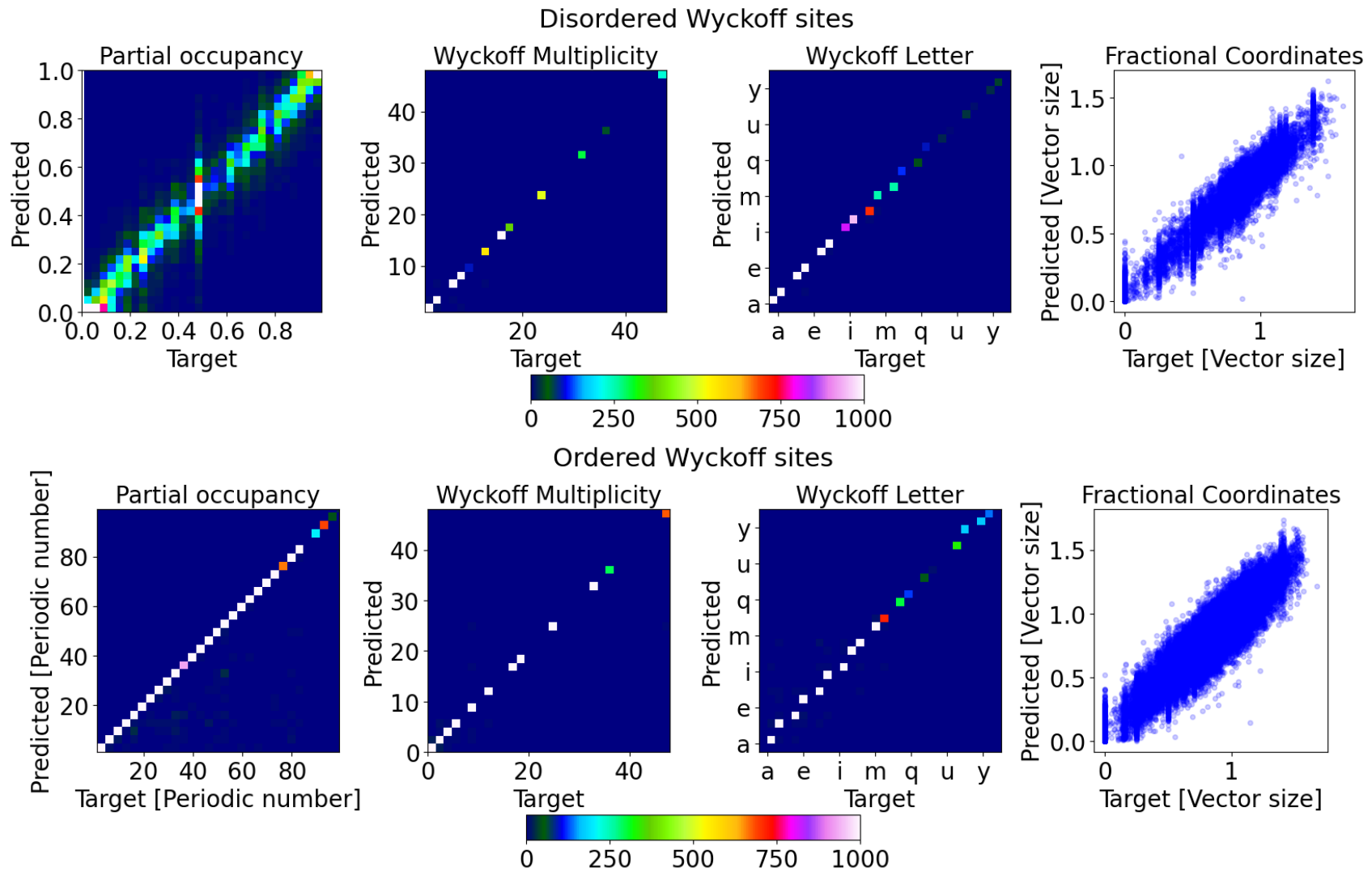}
    \caption{The ordered and disordered Wyckoff site predictions of the test set along the respected color bars. \textbf{Top}: The disordered Wyckoff sites, with partial occupation of atoms along with its other characteristics. \textbf{Bottom}: The ordered Wyckoff sites with an occupation of a single atom along with its other characteristics.}
    \label{fig:wyckoff_site}
\end{figure}

\begin{figure}[t!]
    \centering
    \includegraphics[width=0.4\linewidth]{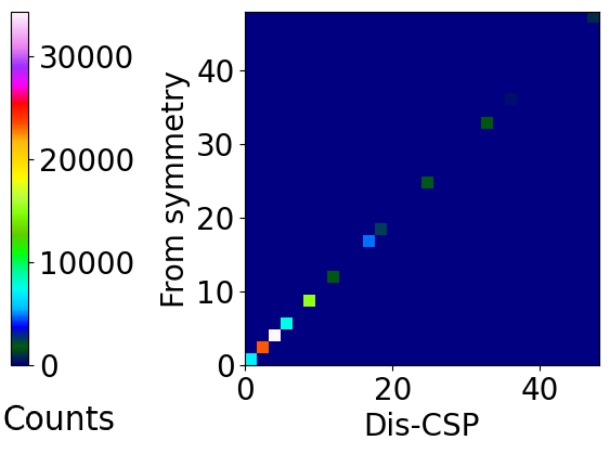}
    \caption{Comparison between the symmetry required and the reconstructed Wyckoff multiplicity of the test set, showing an SMA of 98.4\%.  }
    \label{fig:validation}
\end{figure}

By plotting the reconstructed representation against the target representation for the test set, obvious outliers can be identified, providing insight into the quantitative results presented in \cref{table:result_lattice} and \cref{table:result_wyckoff}. 

In \cref{fig:lattice}, the reconstructed lattice parameters, space group and disordered site indicators are compared to their target values. No obvious outliers are detected in the lattice parameters or space groups, and the Receiver Operating Characteristic (ROC) curve for the disordered site indicator does not indicate significant errors. 

In \cref{fig:wyckoff_site}, the partial occupancy, Wyckoff letter, Wyckoff multiplicity and fractional coordinates are compared to the target values, with a distinction between disordered and ordered Wyckoff sites. For the Wyckoff letter and Wyckoff multiplicity, no obvious outliers are detected. The fractional coordinates exhibit noise around the 1:1 line, consistent with the error rate in \cref{table:result_wyckoff}, and this noise does not differ significantly between the disordered and ordered Wyckoff sites. The partial occupancy also displays noise around the 1:1 line in both cases, although it is more pronounced for the disordered Wyckoff sites. This observation aligns with the quantitative results in \cref{table:result_wyckoff} and suggests that higher accuracy may be required in future training of the VAE model. However, achieving this improvement is challenging due to the high diversity in partial occupancy per Wyckoff site within the dataset. To enhance accuracy, additional filtering strategies may be necessary to refine the dataset, or other training strategies may be needed. 

Moreover, the SMA can be visualized by plotting the reconstructed Wyckoff multipliers, with the one required given the reconstructed Wyckoff letter and space group, as illustrated in \cref{fig:validation}

\newpage
\section{Distribution of properties}
\label{app:Property}
\begin{figure}[h!]
    \centering
    \includegraphics[width=0.7\linewidth]{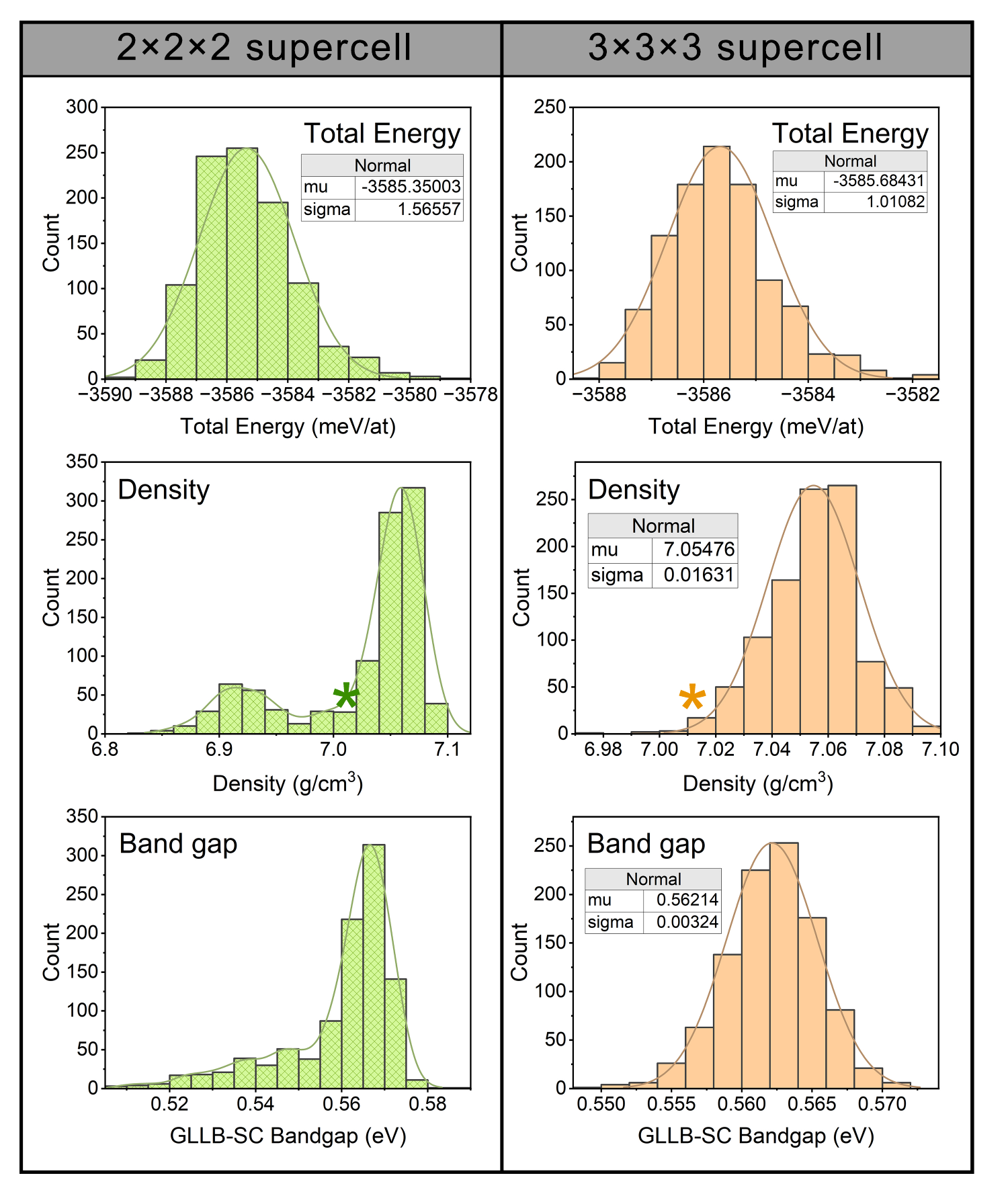}
    \caption{Statistical distributions of selected properties of 1000 virtual cells of site-disordered \ce{AgSbTe_2}. The effect of supercell size (doubled vs. tripled) is compared. In cases where the distribution is approximately normal, the mean (mu) and standard deviation (sigma) are specified. For comparison, the experimental band gap of \ce{AgSbTe_2} is 0.71 eV\cite{abdelghany1996electrical} and the experimental density of \ce{AgSbTe_2} is 7.012 g/cm$^3$ (marked with asterisk)\cite{wu2012reduced}.}
    \label{fig:prop-dist}
\end{figure}
Computed properties of disordered inorganic crystals with partial occupancy are not directly accessible by current first-principles methods or machine-learned force fields. Rather, we estimate these properties by generating a set of \emph{virtual cells}, based on the crystal representation. By using a sufficiently large size of the virtual cell (measured in terms of supercell size) and sufficiently large number of virtual cells used to sample the configurational space, one can approximate the physical properties of the disordered inorganic crystal. Specifically, given a temperature $T$ and using Maxwell-Boltzmann statistics, we can recover the expectation value $\left < P \right >$ of a certain property $P$ from the calculated or predicted values $p_i$ of each virtual cell (of energy $E_i$) in the sample set.
\begin{equation}
    \left < P \right > = \sum_i \frac{p_i e^{\frac{-E_i}{k_{B} T}}}{\sum_j e^{\frac{-E_j}{k_{B} T}}}
\end{equation}
Ignoring the effects of temperature, the properties are better represented as distributions, rather than single number figures. We give an example of the properties of disordered \ce{AgSbTe2} (\cref{fig:prop-dist}), with Ag and Sb being partially occupied at the same Wyckoff site. Total energies and densities are calculated using CHGNET\cite{deng2023chgnet}, and band gaps are predicted using the GLLB-SC model in MEGNet\cite{chen2019graph}. We note that generating virtual cells from the doubled cell creates skewed or bimodal distributions of density and band gap that deviate from those of the tripled supercell, indicating the role of interference from the periodic boundary. 
\section{Conditioned crystal structure generation}
\label{app:Condition}
\begin{figure*}[h!]
    \centering
    \includegraphics[width=1\linewidth]{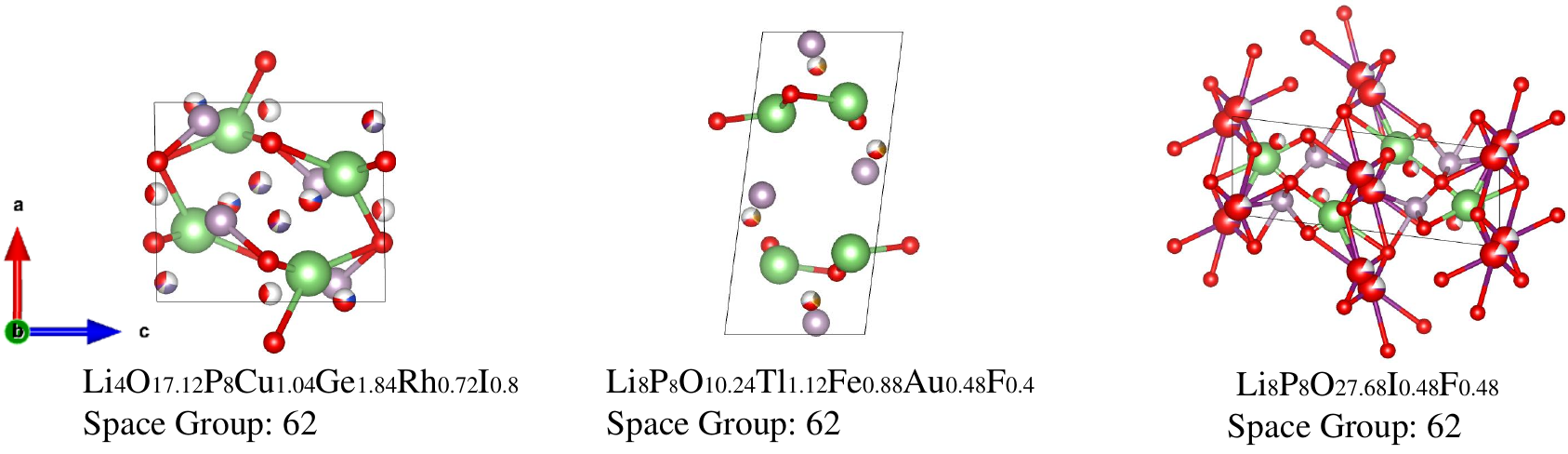}
    \caption{Three generated disordered inorganic crystal structures with space group 62, 1 disordered Wyckoff site and the elements Li, P and O. All crystals are viewed along the b-axis of the crystal.}
    \label{fig:recon_latent}
\end{figure*}

It is possible to condition crystal structure generation based on the framework of Dis-GEN. Thus, it is possible to choose any target value within our representation of the crystal structure. As an example, we utilize gradient descent to explore the latent space, enabling the generation of structures with a space group of 62, a single disordered site, and with the elements \ce{Li}, \ce{P}, and \ce{O} positioned at one of the Wyckoff sites. The gradient descent is utalized by defining a loss function for each condition, as well as use the Adam optimizer\cite{kingma2014adam} to optimize the latent space according to the specified criteria. Generation based on area of the latent space following this conditioning is illustrated in \cref{fig:recon_latent} . If we want to apply further conditions, we use the representation and the loss as described in \cref{sec:model}.

\end{appendices}

\end{document}